\documentclass[epj,twocolumn]{webofc}
\usepackage[mathlines, switch]{lineno}
\usepackage[varg]{txfonts}   % Web of Conferences font
\usepackage{ifthen} 
\newboolean{uprightparticles}
\setboolean{uprightparticles}{false} %Set true for upright particle symbols
\usepackage{xspace} 
\usepackage{upgreek}
\usepackage{amsmath}
\usepackage{xfrac}

\def\lhcb {\mbox{LHCb}\xspace}
\def\ux85 {\mbox{UX85}\xspace}

\def\babar  {\mbox{BaBar}\xspace}
\def\belle  {\mbox{Belle}\xspace}

\def\cleo   {\mbox{CLEO}\xspace}

\ifthenelse{\boolean{uprightparticles}}%
{

 \def\Ppi         {\ensuremath{\uppi}\xspace}

 \def\PDelta      {\ensuremath{\Delta}\xspace}                 
 \def\PXi      {\ensuremath{\Xi}\xspace}                 
 \def\PLambda      {\ensuremath{\Lambda}\xspace}                 
 \def\PSigma      {\ensuremath{\Sigma}\xspace}                 
 \def\POmega      {\ensuremath{\Omega}\xspace}                 
 \def\PUpsilon      {\ensuremath{\Upsilon}\xspace}                 
 
 \def\PA      {\ensuremath{\mathrm{A}}\xspace}                 
 \def\PB      {\ensuremath{\mathrm{B}}\xspace}                 
                  
 \def\PD      {\ensuremath{\mathrm{D}}\xspace}

 \def\PK      {\ensuremath{\mathrm{K}}\xspace}

 \def\PR      {\ensuremath{\mathrm{R}}\xspace}

 \def\Pb      {\ensuremath{\mathrm{b}}\xspace}                 
 \def\Pc      {\ensuremath{\mathrm{c}}\xspace}

 \def\Ph      {\ensuremath{\mathrm{h}}\xspace}                 
 \def\Pi      {\ensuremath{\mathrm{i}}\xspace}

 \def\Ps      {\ensuremath{\mathrm{s}}\xspace}                 
                  
 \def\Pu      {\ensuremath{\mathrm{u}}\xspace}

}
{

 \def\Ppi         {\ensuremath{\pi}\xspace}

 \mathchardef\PDelta="7101
 \mathchardef\PXi="7104
 \mathchardef\PLambda="7103
 \mathchardef\PSigma="7106
 \mathchardef\POmega="710A
 \mathchardef\PUpsilon="7107
 \def\PA      {\ensuremath{A}\xspace}                 
 \def\PB      {\ensuremath{B}\xspace}                 
                  
 \def\PD      {\ensuremath{D}\xspace}

 \def\PK      {\ensuremath{K}\xspace}

 \def\PR      {\ensuremath{R}\xspace}

 \def\Pb      {\ensuremath{b}\xspace}                 
 \def\Pc      {\ensuremath{c}\xspace}

 \def\Ph      {\ensuremath{h}\xspace}                 
 \def\Pi      {\ensuremath{i}\xspace}

 \def\Ps      {\ensuremath{s}\xspace}                 
                  
 \def\Pu      {\ensuremath{u}\xspace}

}

\def\g      {\ensuremath{\Pgamma}\xspace}

\def\uquark    {\ensuremath{\Pu}\xspace}

\def\squark    {\ensuremath{\Ps}\xspace}

\def\cquark    {\ensuremath{\Pc}\xspace}

\def\bquark    {\ensuremath{\Pb}\xspace}

\def\pion  {\ensuremath{\Ppi}\xspace}

\def\pip   {\ensuremath{\pion^+}\xspace}
\def\pim   {\ensuremath{\pion^-}\xspace}

\def\pipm  {\ensuremath{\pion^\pm}\xspace}
\def\pimp  {\ensuremath{\pion^\mp}\xspace}

\def\kaon  {\ensuremath{\PK}\xspace}
  \def\Kbar  {\kern 0.2em\overline{\kern -0.2em \PK}{}\xspace}

\def\Kz    {\ensuremath{\kaon^0}\xspace}
\def\Kzb   {\ensuremath{\Kbar^0}\xspace}
\def\KzKzb {\ensuremath{\Kz \kern -0.16em \Kzb}\xspace}
\def\Kp    {\ensuremath{\kaon^+}\xspace}
\def\Km    {\ensuremath{\kaon^-}\xspace}
\def\Kpm   {\ensuremath{\kaon^\pm}\xspace}
\def\Kmp   {\ensuremath{\kaon^\mp}\xspace}
\def\KpKm  {\ensuremath{\Kp \kern -0.16em \Km}\xspace}
\def\KS    {\ensuremath{\kaon^0_{\rm\scriptscriptstyle S}}\xspace}

  \def\Dbar    {\kern 0.2em\overline{\kern -0.2em \PD}{}\xspace}
\def\D       {\ensuremath{\PD}\xspace}
\def\Db      {\ensuremath{\Dbar}\xspace}
\def\Dz      {\ensuremath{\D^0}\xspace}
\def\Dzb     {\ensuremath{\Dbar^0}\xspace}
\def\DzDzb   {\ensuremath{\Dz {\kern -0.16em \Dzb}}\xspace}
\def\Dp      {\ensuremath{\D^+}\xspace}
\def\Dm      {\ensuremath{\D^-}\xspace}

\def\DpDm    {\ensuremath{\Dp {\kern -0.16em \Dm}}\xspace}

\def\Ds      {\ensuremath{\D_\squark}\xspace}
\def\Dsp     {\ensuremath{\D^+_\squark}\xspace}
\def\Dsm     {\ensuremath{\D^-_\squark}\xspace}
\def\Dspm    {\ensuremath{\D^{\pm}_\squark}\xspace}

\def\B       {\ensuremath{\PB}\xspace}
\def\Bbar    {\ensuremath{\kern 0.18em\overline{\kern -0.18em \PB}{}}\xspace}

\def\Bu      {\ensuremath{\B^+}\xspace}
\def\Bub     {\ensuremath{\B^-}\xspace}
\def\Bp      {\ensuremath{\Bu}\xspace}
\def\Bm      {\ensuremath{\Bub}\xspace}
\def\Bpm     {\ensuremath{\B^\pm}\xspace}

\def\Bs      {\ensuremath{\B^0_\squark}\xspace}
\def\Bsb     {\ensuremath{\Bbar^0_\squark}\xspace}

  \def\Y#1S{\ensuremath{\PUpsilon{(#1S)}}\xspace}% no space before {...}!

\def\Lbar {\ensuremath{\kern 0.1em\overline{\kern -0.1em\PLambda}}\xspace}

\newcommand{\decay}[2]{\ensuremath{#1\!\to #2}\xspace}         % {\Pa}{\Pb \Pc}

\def\to                 {\ensuremath{\rightarrow}\xspace}

\def\CP                {\ensuremath{C\!P}\xspace}

\def\AT#1     {\ensuremath{A_{\mathrm{T}}^{#1}}\xspace}           % 2

\def\C#1      {\ensuremath{\mathcal{C}_{#1}}\xspace}                       % 9
\def\Cp#1     {\ensuremath{\mathcal{C}_{#1}^{'}}\xspace}                    % 7
\def\Ceff#1   {\ensuremath{\mathcal{C}_{#1}^{\mathrm{(eff)}}}\xspace}        % 9  
\def\Cpeff#1  {\ensuremath{\mathcal{C}_{#1}^{'\mathrm{(eff)}}}\xspace}       % 7
\def\Ope#1    {\ensuremath{\mathcal{O}_{#1}}\xspace}                       % 2
\def\Opep#1   {\ensuremath{\mathcal{O}_{#1}^{'}}\xspace}                    % 7

\newcommand{\tev}{\ifthenelse{\boolean{inbibliography}}{\ensuremath{~T\kern -0.05em eV}\xspace}{\ensuremath{\mathrm{\,Te\kern -0.1em V}}\xspace}}
\newcommand{\gev}{\ensuremath{\mathrm{\,Ge\kern -0.1em V}}\xspace}
\newcommand{\mev}{\ensuremath{\mathrm{\,Me\kern -0.1em V}}\xspace}
\newcommand{\kev}{\ensuremath{\mathrm{\,ke\kern -0.1em V}}\xspace}
\newcommand{\ev}{\ensuremath{\mathrm{\,e\kern -0.1em V}}\xspace}
\newcommand{\gevc}{\ensuremath{{\mathrm{\,Ge\kern -0.1em V\!/}c}}\xspace}
\newcommand{\mevc}{\ensuremath{{\mathrm{\,Me\kern -0.1em V\!/}c}}\xspace}
\newcommand{\gevcc}{\ensuremath{{\mathrm{\,Ge\kern -0.1em V\!/}c^2}}\xspace}
\newcommand{\gevgevcccc}{\ensuremath{{\mathrm{\,Ge\kern -0.1em V^2\!/}c^4}}\xspace}
\newcommand{\mevcc}{\ensuremath{{\mathrm{\,Me\kern -0.1em V\!/}c^2}}\xspace}

\def\invfb   {\ensuremath{\mbox{\,fb}^{-1}}\xspace}

\def\fs   {\ensuremath{\rm \,fs}\xspace}

\def\gsim{{~\raise.15em\hbox{$>$}\kern-.85em
          \lower.35em\hbox{$\sim$}~}\xspace}
\def\lsim{{~\raise.15em\hbox{$<$}\kern-.85em
          \lower.35em\hbox{$\sim$}~}\xspace}

\def\tell1  {TELL1\xspace}
\def\ukl1   {UKL1\xspace}

\renewcommand{\g}{$\gamma$\xspace}
\newcommand{\bdk}{\decay{\B}{\D\kaon}}
\newcommand{\bdp}{\decay{\B}{\D\pion}}

\newcommand{\bsdsk}{\decay{\Bs}{\Dspm\Kmp}}
\newcommand{\bbsdsk}{\decay{\boldsymbol{\Bs}}{\boldsymbol{\Dspm\Kmp}}}
\newcommand{\bdh}{\decay{\B}{\D\Ph}}
\newcommand{\bu}{\decay{\bquark}{\uquark}}
\newcommand{\bc}{\decay{\bquark}{\cquark}}
\newcommand{\bpdbkp}{\decay{\Bp}{\Db\Kp}}
\newcommand{\bpdkp}{\decay{\Bp}{\D\Kp}}
\newcommand{\bpdpip}{\decay{\Bp}{\D\pip}}
\newcommand{\bmdkm}{\decay{\Bm}{\D\Km}}
\newcommand{\bmdpim}{\decay{\Bm}{\D\pim}}
\newcommand{\dkp}{\decay{\D}{\Kp\pim}}
\newcommand{\dkppp}{\decay{\D}{\Kp\pim\pip\pim}}
\newcommand{\dkspp}{\decay{\D}{\KS\pip\pim}}
\newcommand{\dkskk}{\decay{\D}{\KS\Kp\Km}}
\newcommand{\bpdhp}{\decay{\Bp}{\D\Ph^+}}
\newcommand{\bmdhm}{\decay{\Bm}{\D\Ph^-}}
\newcommand{\bpmdhpm}{\decay{\Bpm}{\D\Ph^{\pm}}}
\newcommand{\bpmdkpm}{\decay{\Bpm}{\D\Kpm}}
\renewcommand{\tev}{\ensuremath{\mathrm{\,Te\kern -0.1em V}}\xspace}

\woctitle{LHCP 2013}
\begin{document}
\title{Measurement of $\boldsymbol{\gamma}$ from $\boldsymbol{B\rightarrow DK}$ decays at LHCb}
%
% subtitle is optional

\author{Maximilian Schlupp\inst{1}\fnsep\thanks{\email{maximilian.schlupp@tu-dortmund.de}} on behalf of the LHCb Collaboration}

\institute{Technische Universit\"at Dortmund, Experimentelle Physik V, 44227 Dortmund, Germany}

\abstract{
  We report results from the first measurements of the CKM angle \g using \bdk decays 
with the \lhcb experiment. Three well established methods are used to extract the 
\CP observables. The updated measurement of \g in the three-body \Dz Dalitz space 
results in $\gamma = (57\pm 16)^\circ$. 
When combining the observables from all \bdk studies, the best fit value for 
$\gamma \in [0,180]^\circ$ is $\gamma = 67.2^\circ$ with 
$\gamma \in [55.1,79.1]^\circ$ at 68\%CL and $\gamma \in [43.9, 89.5]^\circ$ at 95\%CL. 
This represents the most precise \g values directly measured by a single experiment.\\
Furthermore, a new time-dependent approach using \bsdsk decays is used for the first time 
to measure \CP observables and future 
prospects for \g at \lhcb are given.
}
\maketitle
\section{Introduction}
\label{sec-intro}
The CKM parameter $\gamma = \text{arg}(-V_{ud} V_{ub}^{\ast}/V_{cd} V_{cb}^{\ast})$ 
is the least well measured angle of the Unitarity Triangle. So far, the 
best measurements from single experiments have been performed by the 
\B-factories \babar and 
\belle. The latest results from both experiments are 
$\gamma = (69^{+17}_{-16})^\circ$ \cite{Lees:2013zd} and 
$\gamma = (68^{+15}_{-14})^\circ$ \cite{Trabelsi:2013uj}, respectively.\\
One of the core physics goals of the LHCb experiment is to precisely measure 
the CKM angle \g. This can be 
done by exploiting tree-level processes like \bpmdkpm or \bsdsk, which are sensitive to 
Standard Model (SM) interactions only. In contrast, it is also possible to extract \g from  
loop processes such as two or three-body charmless \B transitions. 
Potential differences in these results 
could indicate new physics contributions. Comparing direct measurements to indirect SM fits 
could also indicate tensions within the SM.\\
Examples of two different approaches to measure \g are described in 
these proceedings. First the more traditional time-independent measurements 
already performed by the \B-factories in section~\ref{sec-bdk} and then 
a new, \lhcb exclusive, time-dependent way in section~\ref{sec-bsdsk}.

\section{Time-Independent measurements using charged $\boldsymbol{\B}$ decays}
\label{sec-bdk}
Measuring \g with charged \bquark-hadron decays one considers 
the interference from \bu and \bc transitions in 
\bdh. Here, \D is either a \Dz or \Dzb and \Ph is a \Kpm or \pipm.\\
The interference is ensured by reconstructing the \D meson in a final 
state common to \Dz and \Dzb, so that the two decay paths \bpdkp 
and \bpdbkp are indistinguishable
\footnote{Charge-conjugation is implied throughout the 
document, if not stated otherwise.}. 
The 
sensitivity on \g is roughly given by the ratio of 
the suppressed over the favoured \B decay amplitude, $r_{\B}$. The 
interference additionally is dependent on the relative strong phase 
difference $\delta_{\B}$ of the two \B amplitudes.\\
There are three established methods to extract \g from these types of 
processes, which depend on the \D final state: the ADS method 
\cite{Atwood:1996ci} using quasi flavour-specific, doubly Cabbibo 
suppressed states (e.g.\ \dkp or \dkppp). 
The \D final states are chosen so that the decay suppressions 
($r_{\B}$ and the \D system equivalent $r_{\D}$) are similar between the two 
interfering \B amplitudes. The \CP asymmetries are therefore expected to 
be large. However, the interference acquires an additional dependence on 
the strong phase difference in the \D meson system, $\delta_{\D}$.\\
The GLW method \cite{Gronau:1990ra,Gronau:1991dp} on the other hand, 
makes use of the \D meson decaying into a \CP eigenstate, where one can eliminate 
the \D system parameters.\\
In the GGSZ method \cite{Giri:2003ty} three-body 
self-conjugate \D final states 
are studied (e.g.\ \dkspp or \dkskk). Performing a Dalitz plot analysis 
of the \D meson decays leads to a good sensitivity on \g.\\
\lhcb results from the three methods are presented in the following sections. 
Additionally, a combination of the various observables from the different 
\B decay modes is shown in section~\ref{sec-comb}, which increases the 
sensitivity on \g beyond the single measurements.
\subsection{ADS/GLW}
\label{sec-GLW}
\begin{figure}
  \centering
  \includegraphics[width=0.47\textwidth,clip]{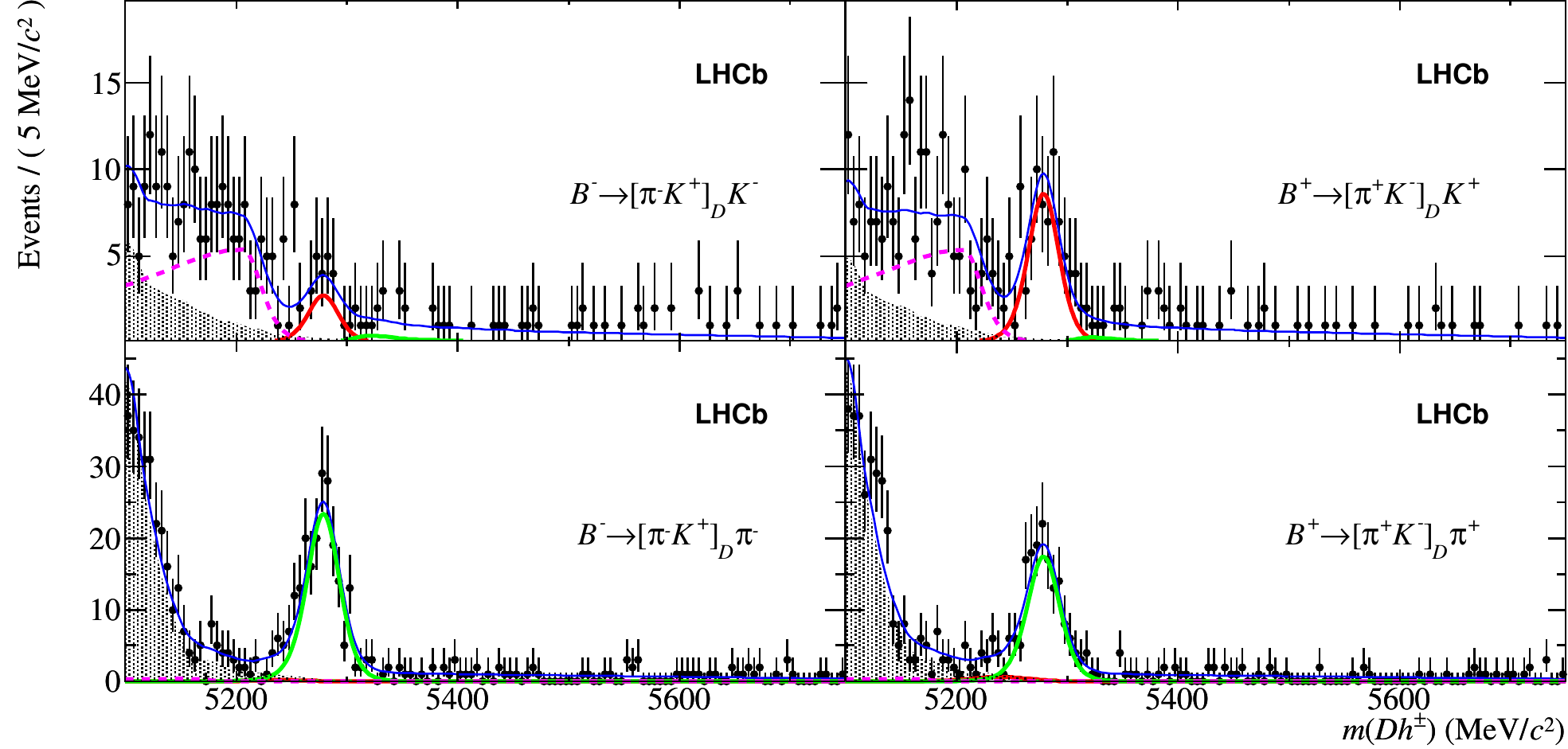}
  %\vspace*{-2mm}
  \caption{Invariant mass distribution of the two-body ADS suppressed modes in 
  $\B\to(\pion\kaon)_{\D}\kaon$ (top) and $\B\to(\pion\kaon)_{\D}\pion$ (bottom).}
  \label{fig:2ADS} 
\end{figure}

\begin{figure}
  %\vspace*{-2mm}
  \centering
  \includegraphics[width=0.47\textwidth,clip]{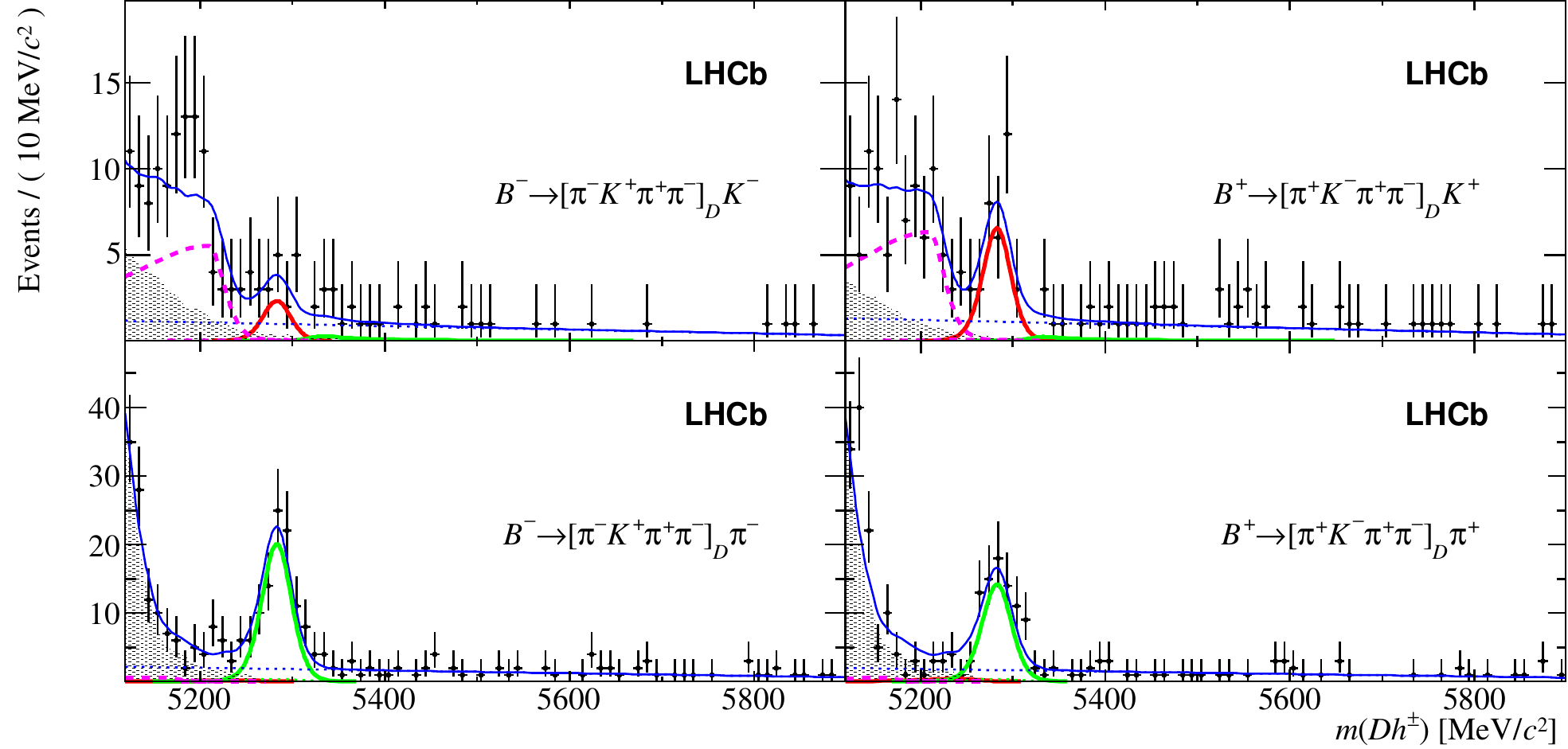}
  %\vspace*{-2mm}
  \caption{Invariant mass distribution of the four-body ADS suppressed modes in 
  $\B\to(\pion\kaon\pion\pion)_{\D}\kaon$ (top) and $
  \B\to(\pion\kaon\pion\pion)_{\D}\pion$ (bottom).}
  \label{fig:4ADS}
\end{figure}

The \lhcb collaboration has performed analyses in \bpdkp and \bpdpip, 
where the \D meson is reconstructed in \Kpm\pimp, \Kp\Km, \pip\pim, \pipm\Kmp, 
and \pipm\Kmp\pip\pim \cite{Aaij:2012kz,Aaij:2013mba} with a dataset 
corresponding to an integrated luminosity of $1\,\invfb$ at $\sqrt{s}=7\tev$. 
The ADS doubly Cabbibo suppressed modes in 
$\B\to(\pion\kaon)_{\D}\kaon$, $\B\to(\pion\kaon\pion\pion)_{\D}\kaon$ and
$\B\to(\pion\kaon\pion\pion)_{\D}\pion$  
are observed for the first time with 
a significance of $>\!\!10\sigma$, $5.1\sigma$ and $>\!\!10\sigma$, respectively. Here 
$(f)_{\D}$ is the abbreviated form for a \D meson decaying into the final state 
$f$, $\D\to f$. The 
respective invariant mass distributions are shown in Figure~\ref{fig:2ADS} 
and \ref{fig:4ADS}.
Using the 
ADS and GLW methods the following \CP observables 
sensitive to \g, $r_{\B}$, $\delta_{\B}$, $r_{\D}$ and $\delta_{\D}$ can be 
measured:
the charge-averaged ratios of \bdk and \bdp
\begin{eqnarray*}
  \PR^{f}_{\kaon/\pi} = \frac{\Gamma(\bmdkm) + \Gamma(\bpdkp)}{\Gamma(\bmdpim) + \Gamma(\bpdpip)} \quad , 
\end{eqnarray*}
where $f$ indicates the \D final state, the charge asymmetries
\begin{eqnarray*}
  \PA^{f}_{\Ph} = \frac{\Gamma(\bmdhm) - \Gamma(\bpdhp)}{\Gamma(\bmdhm) + \Gamma(\bpdhp)} \quad , 
\end{eqnarray*}
and the non charge-averaged ratio of suppressed and favoured \D final state
\begin{eqnarray*}
  \PR^{\pm}_{\Ph} = \frac{\Gamma(\bpmdhpm)_{\text{sup}}}{\Gamma(\bpmdhpm)} \quad . 
\end{eqnarray*}
The resulting values can be found in the refs. \cite{Aaij:2012kz,Aaij:2013mba} 
and serve as inputs for the combined \g measurement in section~\ref{sec-comb}. 
Furthermore, direct \CP violation in \bpmdkpm is observed with a total significance of 
$5.8\sigma$.

\subsection{GGSZ}
\label{sec-GGSZ}
\begin{figure}
  \centering
  \includegraphics[width=0.47\textwidth]{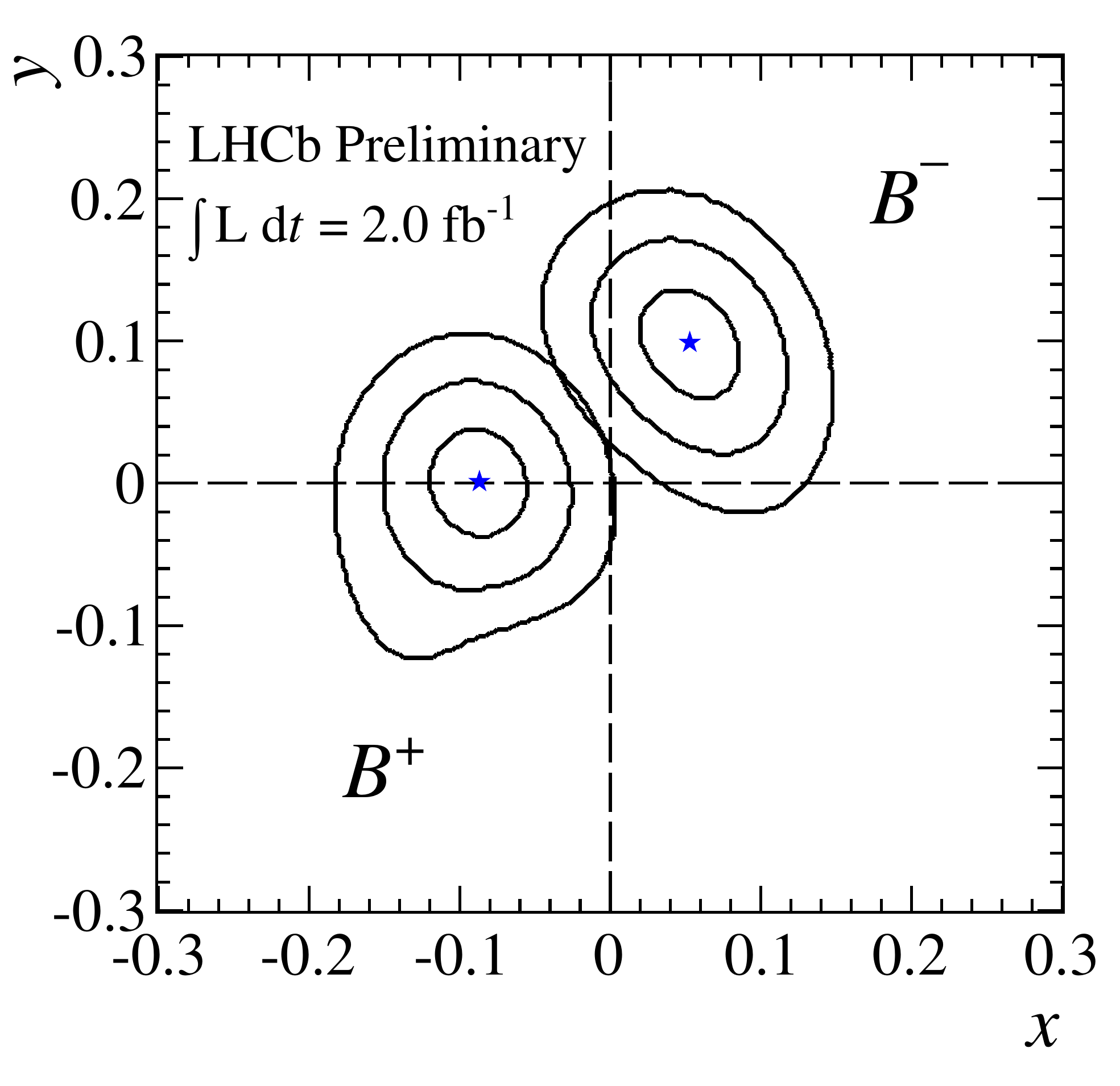}
  %\vspace*{-2mm}
  \caption{Best fit values (stars) and $1\sigma$, $2\sigma$ and $3\sigma$ confidence 
  intervals (contours) in the ($x$,$y$) plane using the statistical uncertainties and 
  correlations only.}
  \label{fig:ggszxy}
\end{figure}

The GGSZ method exploits the three-body $\D\to\KS\Ph^+\Ph^-$ Dalitz space 
in \bpmdkpm decays to extract the \CP observables 
$x_{\pm} = r_{\B}\cos(\delta_{\B}\pm\gamma)$ 
and $y_{\pm} = r_{\B}\sin(\delta_{\B}\pm\gamma)$. 
Due to the rich resonance structure of the \D decays, this method has proven to be 
most sensitive one at the \B-factories.
We report the model-independent measurement using a dataset corresponding to 
$2\,\invfb$ of integrated luminosity with a centre of mass energy of $\sqrt{s}=8\tev$ 
by the LHCb Collaboration \cite{LHCb-CONF-2013-004}, which is the successor of the 
$1\,\invfb$ publication \cite{Aaij:2012hu} at $\sqrt{s}=7\tev$. 
The variation of the strong phase 
difference $\delta_{\D}$ in bins of the $\D\to\KS\Ph^+\Ph^-$ Dalitz plot is taken 
as an external input from the \cleo collaboration. 
The resulting numbers for the \CP violation parameters $x_{\pm}$ and $y_{\pm}$ are 
illustrated in Figure~\ref{fig:ggszxy} for $2\,\invfb$, where the combined 
$3\,\invfb$ values are:
\begin{linenomath*}
\begin{align*}
\hspace*{-2mm}
\langle x_+ \rangle &= (-8.9\pm3.1)\times 10^{-2}{,}\ \   
\langle x_- \rangle = (3.5\pm2.9)\times 10^{-2}\\
\langle y_+ \rangle &= (0.1\pm3.7)\times 10^{-2}{,}\ \ \ \ \,  
\langle y_- \rangle = (7.9\pm3.8)\times 10^{-2}.
\end{align*}
\end{linenomath*}
The dominant systematic uncertainties are coming from the assumption of no interference 
in the control channel and the external hadronic input parameters. However, the results are 
limited statistically.
The underlying physics parameters are extracted using a frequentist approach resulting 
in $\gamma = (57\pm16)^\circ$, $r_{\B} = (8.8^{+2.3}_{-2.4})\times 10^{-2}$ and 
$\delta_{\B} = (124^{+15}_{-17})^\circ$. This results competes with the methodically 
equivalent \belle measurement \cite{Aihara:2012aw} of 
$\gamma = (77.4^{+15.1}_{-14.9}\pm 4.1 \pm 4.3)^\circ$ for the current world's most precise 
single direct measurement of \g. 
\subsection{Combination}
\label{sec-comb}
\begin{figure}
  \vspace*{-2mm}
  \centering
  \includegraphics[width=0.47\textwidth]{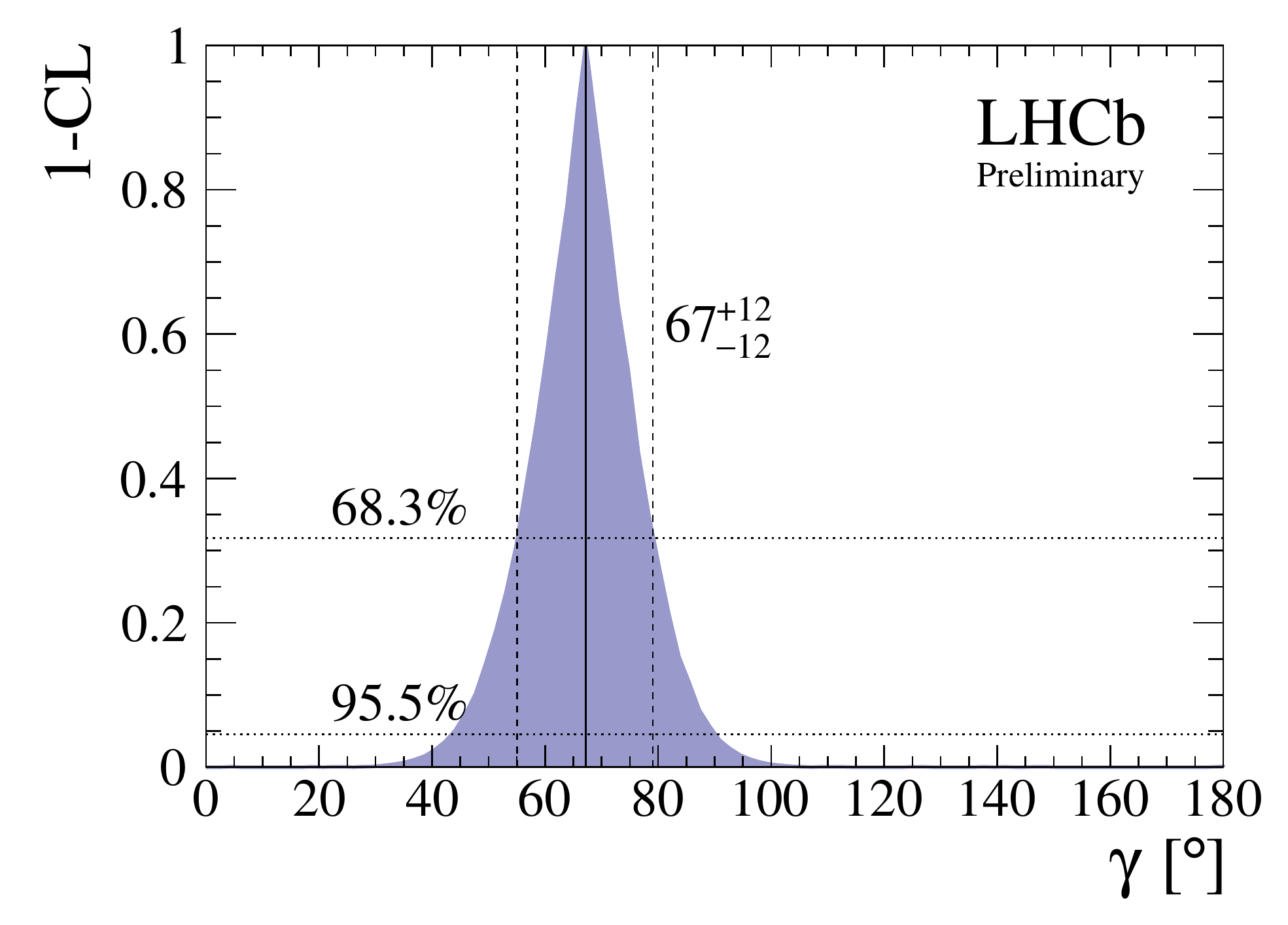}
  %\vspace*{-2mm}
  \caption{$1-\text{CL}$ curve for \g from the combined ADS/GLW $1\,\invfb$ and GGSZ 
  $3\,\invfb$ measurements. The $1\sigma$ and $2\sigma$ confidence interval can be 
  read off at the intersections of the blue curve with the dotted lines labelled 
  $68.3\,\%$ and $95.5\,\%$, respectively.}
  \label{fig:combG}
\end{figure}

\begin{figure}
  \vspace*{-2mm}
  \centering
  \includegraphics[width=0.47\textwidth]{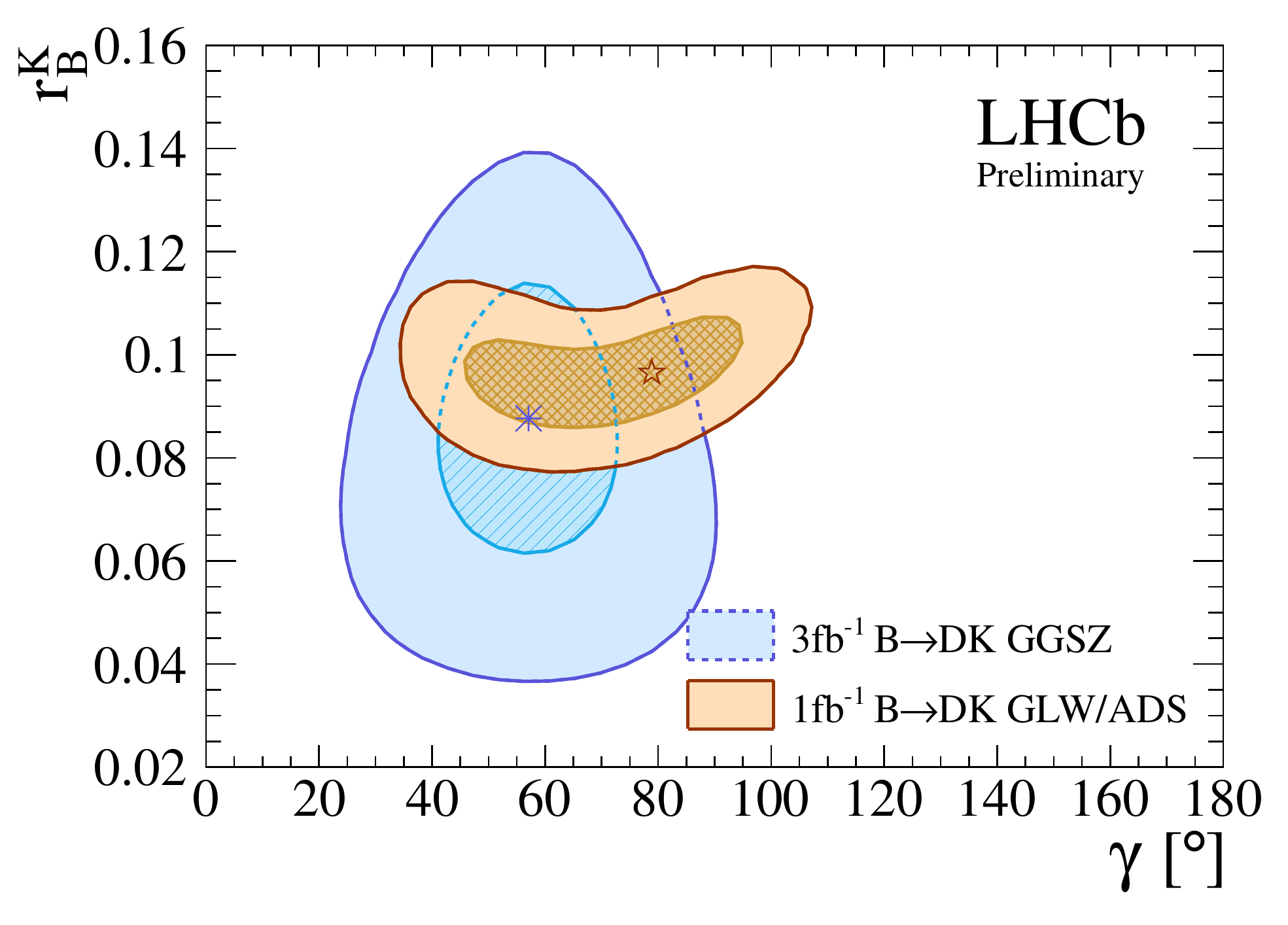}
  %\vspace*{-2mm}
  \caption{Best-fit values (markers) and contours where the difference in log-likelihood 
  corresponds to $1\sigma$ and $2\sigma$. The $3\,\invfb$ GGSZ and $1\,\invfb$ ADS/GLW analyses are shown separately in blue and orange.}
  \label{fig:comb2d}
\end{figure}

To reach the best possible sensitivity on \g the observables from the ADS, GLW and GGSZ 
analyses, the amplitudes and ratios from section~\ref{sec-GLW} and the combined 
$3\,\invfb$ \CP observables from section~\ref{sec-GGSZ}, are evaluated at the same time 
for the \bdk transitions. 
Additionally, inputs from the \cleo collaboration
\cite{Lowery:2009id} and the Heavy Flavour Averaging Group (HFAG)
\cite{Amhis:2012bh} have been used to constrain the hadronic parameters of the \D system 
and the effect of direct \CP violation in \D decays, respectively.
A likelihood is constructed from the input measurements as follows:
\begin{eqnarray*}
 \mathcal L(\vec{\alpha}) = \prod_i{\xi_i(\vec{A}_i^{\text{obs}}|\vec{\alpha})}\quad ,
\end{eqnarray*}
where $i$ denotes the different measurements, $\vec{A}_i^{\text{obs}}$ the observables, 
$\xi_i$ the probability density functions (PDFs) of the observables $\vec{A}_i$ 
and $\vec{\alpha}$ is the set of parameters (\g, $r_{\B}$, etc.). For most of the PDFs 
$\xi_i$ a multidimensional Gaussian is assumed taking correlations into account. 
Whenever highly non-Gaussian behaviour is present, $\xi_i$ is replaced by the 
experimental likelihood.\\
The confidence intervals are calculated using a frequentist method. Its coverage 
is not guaranteed from first principles, so the coverage is tested. It is found that the 
coverage is almost correct so that the results are scaled according to the small 
differences. Additionally, the confidence intervals are cross-checked 
and found to be consistent 
with a method inspired by Berger and Boos \cite{BergerBoos}. In this method the 
values of the nuisance parameters are sampled from a uniform distribution covering 
a multidimensional confidence belt $C_{\beta}$, instead of fixing the nuisance 
parameters to their best-fit values. $C_{\beta}$ is chosen such that the 
corresponding corrections to the p-value are negligible. 
For more details on the inputs, the statistical procedures and the validation 
of the results, see \cite{Aaij:2013zfa, LHCb-CONF-2013-006, BergerBoos}.
The best fit values and confidence intervals for \g, $r_{\B}$ and $\delta_{\B}$ are listed 
in Table~\ref{tab:com}, all values are modulo $180^\circ$.
\begin{table}
\centering
\caption{Best-fit values and confidence intervals for \g, $r_{\B}$ and $\delta_{\B}$ from the combination of the \bdk measurements.}
\label{tab:com} 
\begin{tabular}{ll}
\hline
quantity & \D\hspace*{-2px}\kaon combination  \\\hline
\g            & $67.2^\circ$  \\
68\% CL       & $[55.1,79.1]^\circ$ \\
95\% CL       & $[43.9,89.5]^\circ$ \\\hline
$r_{\B}$      & $114.3^\circ$  \\
68\% CL       & $[101.3,126.3]^\circ$ \\
95\% CL       & $[88.7,136.3]^\circ$ \\\hline
$\delta_{\B}$ & $0.0923$  \\
68\% CL       & $[0.0843,0.1001]$ \\
95\% CL       & $[0.0762,0.1075]$ \\\hline
\end{tabular}
\end{table}
The $1-\text{CL}$ curve for \g and the two-dimensional likelihood projection 
for \g and $r_{\B}$ 
are shown in Figure~\ref{fig:combG} and ~\ref{fig:comb2d}, respectively. 
The 68\% CL interval for \g can be translated to $\gamma = (67\pm12)^\circ$. 
This 
preliminary result has a lower uncertainty compared to the latest results from 
\babar \cite{Lees:2013zd} and \belle \cite{Trabelsi:2013uj}.%\\
\section{Time-dependent measurement in \bbsdsk decays}
\label{sec-bsdsk}
\begin{figure}
  %\vspace*{-2mm}
  \centering
  \includegraphics[width=0.47\textwidth]{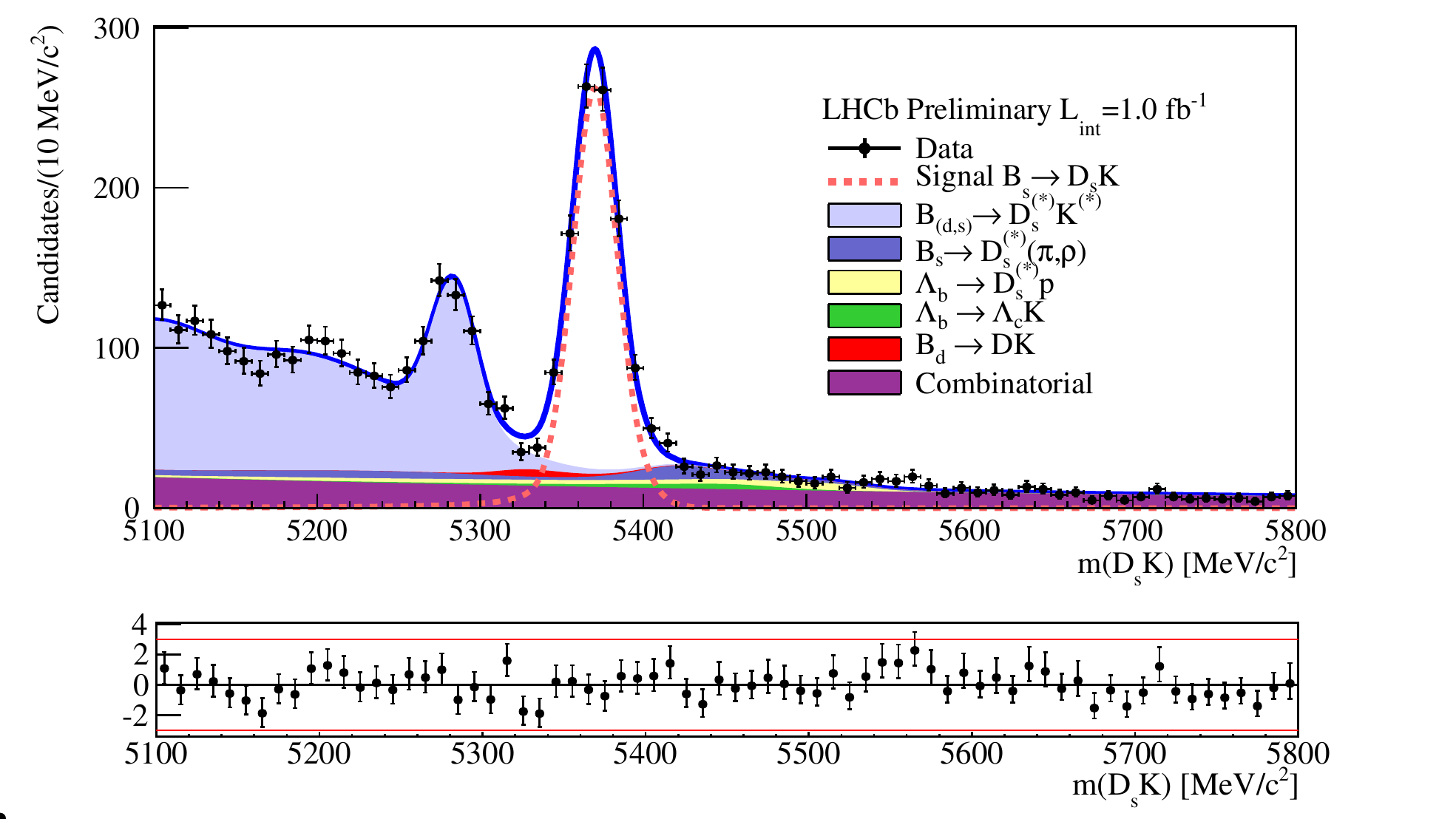}
  %\vspace*{-2mm}
  \caption{Invariant mass distribution of \Bs candidates together with the signal and 
  background components and the full fit. Below the corresponding pulls are shown.}
  \label{fig:massFit}
\end{figure}

\begin{figure}
  %\vspace*{-2mm}
  \centering
  \includegraphics[width=0.47\textwidth]{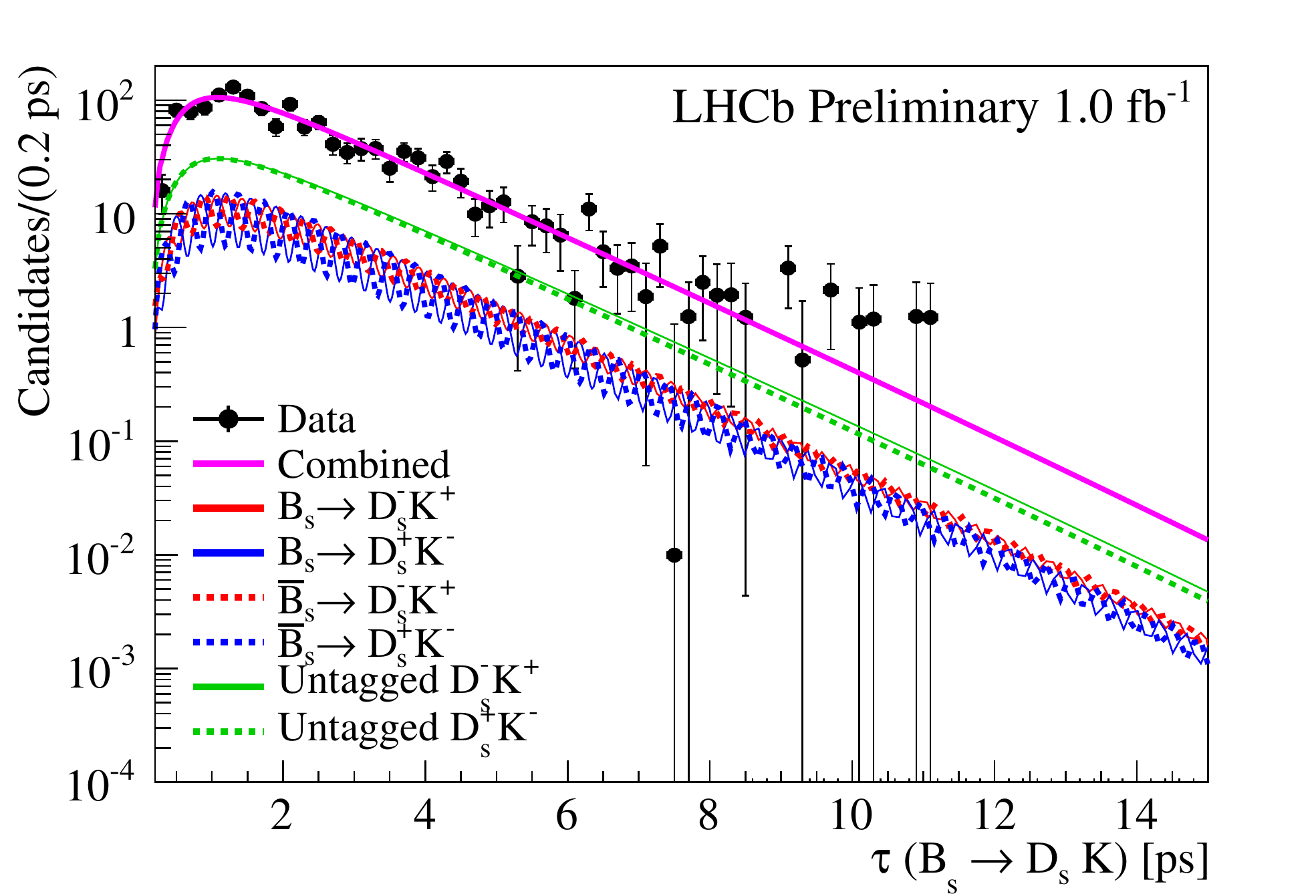}
  %\vspace*{-2mm}
  \caption{Fit to the weighted decay time distribution, showing all 
  fit components separately.}
  \label{fig:dectimeFit}
\end{figure}

A different approach to extract \g is to use 
neutral \B mesons and perform a time-dependent measurement of the 
\CP parameters. This 
can be done using tree-level \bsdsk decays. The sensitivity to \g arises 
from the interference of both \B mesons, \Bs and \Bsb, 
decaying into the same final state: \Dsp\Km or \Dsm\Kp. Note that the \Ds 
final states are not of major importance in this method. Each decay amplitude 
is roughly of the same order of magnitude, thus the expected interference is 
large $r_{\B}^{\Ds\kaon} = 0.37$.\\
In order to resolve the \Bs oscillations, a good time resolution 
is mandatory. For the analysis of \bsdsk decays at \lhcb \cite{LHCb-CONF-2012-029} 
it is determined from Monte Carlo (MC) simulations. The 
difference of the reconstructed and the true decay time is 
fitted with a resolution model, which is the sum of three Gaussians.
To account for differences in data and simulations we scale the Gaussian's widths 
according to \Bs\to\Ds\pion MC and a data 
sample of "fake" \Bs constructed from prompt \Ds mesons which are combined with a 
random \pion. We assume that the differences between \bsdsk and the 
control channel \Bs\to\Ds\pion are negligible for the relevant quantities.
The resulting effective time resolution is estimated as 
$\sigma_t \approx 50\fs$. Another crucial part is the determination of 
the time acceptance, which is also obtained from MC. 
The invariant mass distribution of the \Bs candidates is fitted using an unbinned maximum 
likelihood method in order to get weights, which separate signal from background components.
The full mass-fit is shown in Figure~\ref{fig:massFit}.
%An \emph{sFit} \cite{sFit} is performed to the invariant mass distribution of the 
%\Bs candidates in order to get weights which separate signal from background components.
The weighted decay time distribution is then fitted using the 
\emph{sFit} \cite{sFit} 
technique, where the fit determines the corresponding \CP observables.
The resulting values can be found in \cite{LHCb-CONF-2012-029} and the decay 
time fit is shown in Figure~\ref{fig:dectimeFit}.
The weighing procedure is cross-checked with a conventional 2-dimensional 
fit in the invariant mass and decay time.\\
It is found that correlations within the systematics have a non-negligible effect on 
extracting the actual \CP parameters $\gamma + \beta_s$, where $\beta_s$ is the \Bs 
mixing phase. Measuring the \CP parameters marks the first important step 
towards a time-dependent estimation of \g from \bsdsk decays.
\section{Conclusions and prospects}
\label{sec-conc}
We reported several measurements of \g with the \lhcb experiment. Up to now, the 
GGSZ analysis is the most sensitive single measurement 
of $\gamma = (57\pm16)^\circ$ using the full combined $3\,\invfb$ \lhcb dataset. Exploiting 
the ADS/GLW method on $1\,\invfb$ of \lhcb data
in \bdh with two- and four-body \D decays leads to the observations of the 
corresponding suppressed ADS modes with significances greater than $5\sigma$. 
Furthermore, \CP observables are provided by the analyses from which \g can be extracted.\\
Combining all \CP observables from the \bdk measurements the resulting \lhcb result is 
$\gamma = (67\pm12)^\circ$, which is more precise than recent \babar \cite{Lees:2013zd} 
and \belle \cite{Trabelsi:2013uj} results.
Further improvements are expected with the analyses updated to the full available dataset. 
When more channels, which were not discussed throughout these proceedings are 
analysed with the current or with a future dataset, the sensitivity on \g will 
increase by including these to the combined measurement. Then \lhcb will be able to 
compare \g estimations from tree-level and loop-level processes. \\
In the future we expect to decrease the uncertainty on \g to 
$\delta \gamma \sim \mathcal O(1^\circ)$ \cite{Bediaga:2012py} using 
a dataset of $50\,\invfb$ and combining different 
decay channels. This dataset is planed to be recorded within the coming decade. 
\begin{acknowledgement}
I would like to thank the organisers of the LHCP 2013 
for the possibility to participate in 
this excellent conference.\\
This work is financed by the german Federal Ministry of Education and Research 
(BMBF).
\end{acknowledgement}

\bibliography{LHCP2013_proceedings}
\end{document}